\documentclass[multphys,vecphys]{svmult}

\usepackage{makeidx}     
\usepackage{graphicx}    
\usepackage{multicol}    

\makeindex             

\begin{document}

\title*{How Common are Engines in Ib/c Supernovae?}
\author{Edo Berger}
\institute{Division of Physics, Mathematics and Astronomy, 105-24,
California Institute of Technology, Pasadena, CA 91125
\texttt{ejb@astro.caltech.edu}}
\maketitle

\begin{abstract}
The association of $\gamma$-ray bursts (GRBs) and core-collapse
supernovae (SNe) of Type Ib and Ic was motivated by the detection of
SN\,1998bw in the error box of GRB\,980425 and the now-secure
identification of a SN\,1998bw-like event in the cosmological
GRB\,030329.  The bright radio emission from SN\,1998bw indicated that
it possessed some of the unique attributes expected of GRBs, namely a
large reservoir of energy in (mildly) relativistic ejecta and variable
energy input.  Here we discuss the results of a systematic program of
radio observations of most reported Type Ib/c SNe accessible to the
Very Large Array, designed to determine the fraction of Type Ib/c SNe
driven by an engine.  We conclude that: (i) the incidence of such
events is low, $<3\%$, and (ii) there appears to be a clear dichotomy
between the majority of hydrodynamic explosions (SNe) and
engine-driven explosions (GRBs).
\end{abstract}

\section{Hydrodynamic vs.~Engine Driven Explosions}
\label{sec:1}

Stellar explosions can be characterized by their kinetic energy,
$E_K$, and the mass of the ejecta, $M_{\rm ej}$.  Equivalently one may
consider $E_K$ and the mean initial speed of ejecta, $v_0$, or the
Lorentz factor, $\Gamma_0= [1-\beta_0^2]^{-1/2}$, where
$\beta_0=v_0/c$.  In this context, supernovae (SNe) and $\gamma$-ray
bursts (GRBs), are distinguished by their ejecta velocities: $v_0\sim
10^4$ km s$^{-1}$ as inferred from optical absorption features
(e.g.~\cite{fil97}), and $\Gamma_0>100$, inferred from the non-thermal
prompt emission \cite{goo86,pac86}, respectively.

In the conventional interpretation, $M_{\rm ej}$ for SNe is large
because $E_K$ is derived from the (essentially) symmetrical
collapse of the core and the energy thus couples to all the mass left
after the formation of the compact object.  

GRB models, on the other hand, appeal to an engine --- a stellar mass
black hole, which accretes matter on many dynamical timescales and
powers relativistic jets (the so-called collapsar model;
\cite{woo93}).  Observationally, this model is supported by the
complex temporal profiles and long duration of GRBs, their high Lorentz
factors, a high degree of asymmetry \cite{fks+01}, and episodes of
energy injection.

\section{SN\,1998bw: An Engine Driven Supernova}
\label{sec:bw}

The unusual SN\,1998bw shares some of the unique attributes expected
of GRBs.  This Type Ic SN coincided in time and position with
GRB\,980425 \cite{gvv+98}, for which the inferred isotropic energy in
$\gamma$-rays was only $8\times 10^{47}$ erg \cite{paa+00}, three
to six orders of magnitude fainter than typical GRBs.  More
importantly, SN\,1998bw exhibited unusually bright radio emission
indicating about $10^{50}$ erg of mildly relativistic ejecta as 
well as variable energy input \cite{lc99}.
To date these features have not been seen in any other nearby SN.  
Thus, the empirical data strongly favor an engine in SN\,1998bw.

Two scenarios for the origin of SN\,1998bw and its relation to GRBs
have been proposed: (i) GRB\,980425 may have been a typical burst but
viewed well away from the jet axis (hereafter, the off-axis model),
and (ii) SN\,1998bw represents a different class of SNe.

A powerful discriminant between these two scenarios is the expected
rate of SN\,1998bw-like events.  In the off-axis model, the fraction
of Type Ib/c SNe that are powered by a central engine is linked to the
mean beaming factor of GRBs, $f_b$ (e.g.~\cite{fks+01}); a recent
estimate is $\langle f_b^{-1}\rangle\sim 500$ \cite{fks+01}.  Coupled
with an estimated local GRB rate of $\sim 0.5$ Gpc$^{-3}$ yr$^{-1}$
\cite{sch01} compared to a Type Ib/c SN rate of $\sim 4.8\times
10^{4}$ Gpc$^{-3}$ yr$^{-1}$ \cite{mcp+98,cet99,frp+99}, we expect
that $\sim 0.5\%$ of Type Ib/c SNe will be similar to SN\,1998bw.

On the other hand, if SN\,1998bw is not an off-axis burst, then the
rate of similar events has to be assessed independent of the GRB rate.
In this context, Norris (2002)~\cite{nor02} has argued that of the
$1429$ long-duration BATSE bursts, about 90 events possess similar
high-energy attributes as that of GRB\,980425.  This number
corresponds to about $25\%$ of Type Ib/c SNe within 100 Mpc.

\section{A VLA Survey of Type Ib/c Supernovae}
\label{sec:vla} 

Our basic hypothesis is that (mildly) relativistic ejecta are best
probed by radio observations, as was demonstrated in the case of
SN\,1998bw.  To this end we began a program of observing most reported
Type Ib/c SNe with the Very Large Array in late 1999 \cite{bkf+03}.

\begin{figure}
\centering
\includegraphics[height=9.5cm]{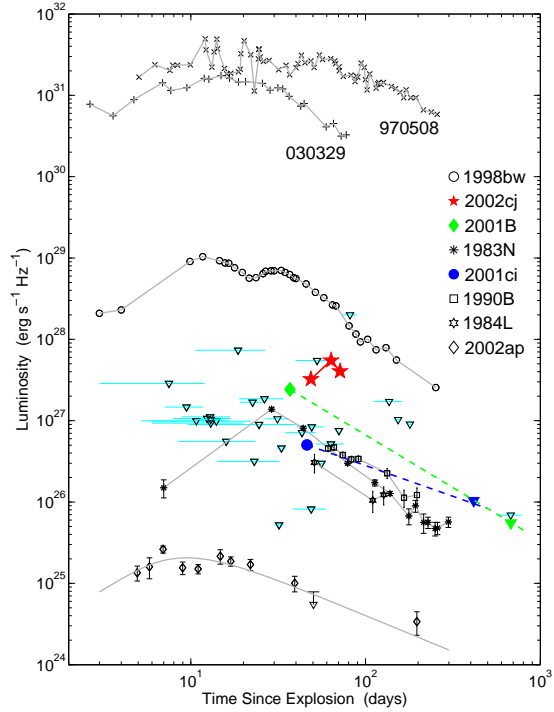}
\caption{Radio lightcurves of Type Ib/c SNe detected in this survey
and from the literature, as well as upper limits for the
non-detections (Ref.~\cite{bkf+03} and references therein).  We also
include the radio lightcurves of GRB\,970508  \cite{fwk00} and
GRB\,030329 \cite{bkp+03}.  The uncertainty in time for the
non-detections represents the uncertain time of explosion.}
\label{fig:lcs}
\end{figure}

Figure~\ref{fig:lcs} provides a succinct summary of the radio
lightcurves and upper limits.  Three strong conclusions can be
drawn from this Figure.  First, SNe as bright as SN\,1998bw are rare;
we find a limit of $<3\%$ from our survey.  Second, there is
significant dispersion in the luminosities of Type Ib/c SNe.  Finally,
the radio emission from SNe (including SN\,1998bw) is orders of
magnitude dimmer than that of GRB afterglows.

\subsection{Expansion Velocities and Energetics}
\label{sec:vels}

In the framework of synchrotron self-absorption, the peak time and
peak luminosity directly measure the mean expansion speed of the
fastest ejecta \cite{che98}.  We infer velocities ranging from $v\sim
10^4$ to $10^5$ km s$^{-1}$ based on our detections and upper limits
\cite{bkf+03}.

We also find that the ejecta giving rise to the radio emission from 
SNe for which detailed
information is available (SN\,1984L and SN\,2002ap) can be produced by
a hydrodynamic explosion \cite{bkc02,bkf+03}.  In fact, the estimated
energies from the hydrodynamic models \cite{che82,mm99,inn+00} exceed
those inferred from the radio observations by up to two orders of
magnitude.  This may indicate that the total kinetic energies have
been over-estimated, possibly as a result of neglecting a mild
asymmetry.

We therefore conclude that none of the SNe observed in our 
survey and in the past clearly exhibits the unique characteristics of 
SN\,1998bw: a significant excess of energy in mildly relativistic 
ejecta.

\section{A Comparison to $\gamma$-Ray Burst Afterglows}
\label{sec:comp}

\begin{figure}
\centering
\includegraphics[height=8.8cm]{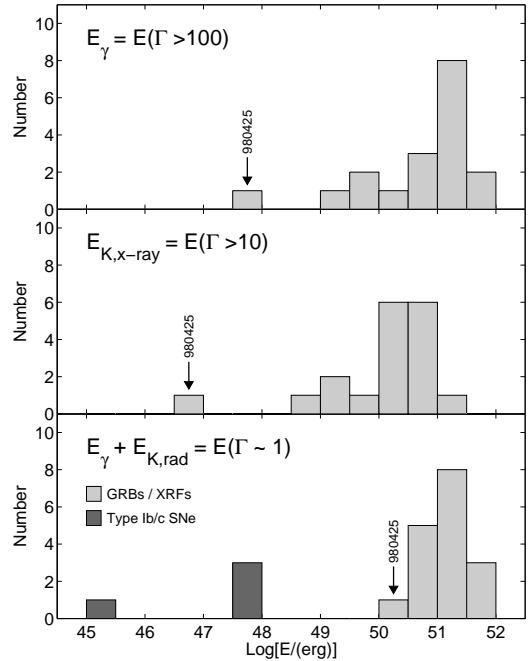}
\caption{Histograms of the beaming-corrected $\gamma$-ray energy
\cite{bfk03}, $E_\gamma$, the kinetic energy inferred from X-rays at
$t=10$ hr \cite{bkf03}, $E_{K,X}$, and total relativistic energy,
$E_\gamma+E_{K}$, where $E_K$ is the beaming-corrected kinetic energy
inferred from the broad-band afterglows of GRBs \cite{lc99,pk02} and
radio observations of SNe.  The wider dispersion in $E_\gamma$ and
$E_{K,X}$ compared to the total energy indicates that engines in
cosmic explosions produce approximately the same quantity of energy,
thus pointing to a common origin, but the ultra-relativistic output of
these engines varies widely.  In Type Ib/c SNe, on the other hand, the
total explosive yield in fast ejecta (typically $\sim 0.3c$) is
significantly lower.}
\label{fig:hist1}
\end{figure}

From Figures~\ref{fig:lcs} and \ref{fig:hist1} it is clear that the
radio lightcurves of GRB afterglows and SNe are dramatically
different.  This has significant implications, namely {\it none} of
the Type Ib/c SNe presented in Figure~\ref{fig:lcs} could have given
rise to a typical $\gamma$-ray burst.  However, SN\, 1998bw is unique
in both samples: it is fainter than typical radio afterglows of GRBs
but much brighter than Type Ib/c SNe (Figure~\ref{fig:hist1}).

\subsection{Hypernovae}

The discovery of broad optical lines and large explosive
energy release, $>few$ FOE, in SN\,1998bw prompted some astronomers to
use the designation ``hypernovae'' for SN\,1998bw-like SNe.
Unfortunately, this designation is not well defined, and has been
applied liberally in recent years.

In our framework the critical distinction between an ordinary
supernova and a GRB explosion is relativistic ejecta carrying a
considerable amount of energy.  Such ejecta are simply not traced by
optical spectroscopy.  This reasoning is best supported by the fact
that the energy carried by the fast ejecta in SN\,1998bw and
SN\,2002ap \cite{bkc02} differ by four orders of magnitude even
though both exhibit broad spectral features at early times and both 
have been called hypernovae.

\section{Conclusions}
\label{sec:conc}

We end with the following conclusions.  First, radio observations
provide a robust way of measuring the quantity of energy associated
with high velocity ejecta.  This allows us to clearly discriminate
between engine-driven SNe such as SN\,1998bw and ordinary SNe, powered
by a hydrodynamic explosion, such as SN\,2002ap \cite{bkc02}.  Second,
at least $97\%$ of local Type Ib/c SNe are not powered by engines and
furthermore have a total explosive yield of only $10^{48}$ erg in fast
ejecta.  As summarized in Figure~\ref{fig:hist1}, this indicates that
there is a clear dichotomy between Type Ib/c SNe and cosmic,
engine-driven explosions.



\printindex

\end{document}